\documentclass[a4paper,onecolumn,showkeys,floatfix,aps,pre,preprint,groupedaddress]{revtex4-1}
\usepackage{color}
\usepackage{amsmath,amssymb}
\usepackage{graphics,graphicx}
\usepackage{dcolumn,bm}
\usepackage{psfrag}
\usepackage{enumerate}

\begin{document}

\title{Optimization strategies of human mobility during the COVID-19 pandemic: A review}
\author{Soumyajyoti Biswas${}^{1}$}
\email{soumyajyoti.b@srmap.edu.in}
\author{Amit Kr Mandal${}^{2}$}
\email{amitmandal.nitdgp@gmail.com}

\affiliation{
${}^1$ Department of Physics, SRM University, AP - Amaravati 522502, Andhra Pradesh, India \\
${}^2$ Department of Computer Science and Engineering, SRM University - AP, Andhra Pradesh - 522502, India 
}

\date{\today}

\begin{abstract}
The impact of the ongoing COVID-19 pandemic is being felt in all spheres of our lives -- cutting across the boundaries of nation, 
wealth, religions or race. From the time of the first detection of infection among the public, the virus spread though almost all
the countries in the world in a short period of time. With humans as the carrier of the virus, the spreading process necessarily 
depends on the their mobility after being infected. Not only in the primary spreading process, but also in the subsequent spreading of the mutant variants, 
human mobility plays a central role in the dynamics. Therefore, on one hand travel restrictions of varying degree were imposed and are
still being imposed, by various countries both nationally and internationally. On the other hand, these restrictions have severe 
fall outs in businesses and livelihood in general. Therefore, it is an optimization process, exercised on a global scale, with multiple
changing variables. Here we review the techniques and their effects on optimization or proposed optimizations of human mobility in different 
scales, carried out by data driven, machine learning and model approaches.     
\end{abstract}


\maketitle


\section{Introduction}
In the lack of any known treatment protocol or that of a cure, one of earliest responses of the outbreak of 
SARS-CoV-2 or the COVID-19 disease \cite{who} was to establish a boundary around the epicenter of the outbreak (Hubei province in China)
-- a cordon sanitaire -- on January 23, 2020 to prevent the infection from spreading. It had, nevertheless, spread out, triggering similar responses 
from various other countries at varying degree of duration and scale of restrictions (see e.g., \cite{unwto}). Many such restrictions exist till date, while
some of it were lifted either for a short or a longer time. 

Indeed, cordon sanitaire is an old technique of infectious disease containment. The use of the phrase goes back to 1821, when 30000
French troops were deployed by Duke de Richelieu apparently to prevent yellow fever to spread from Barcelona to France \cite{taylor}, 
but its first documented use dates even further back to 1523 in Malta \cite{malta}. With a varying degree of successes in the past, 
the scale of its implementation has never been larger than the current one -- affecting almost the entire population on the planet.
While it is still early to discuss the full impact of such restrictions on different spheres of the society, it is 
possible to assess some of the impacts of the restrictions on spreading of the disease, on early economic fallout and the burdens
placed on the health infrastructures. 

It is possible to place the question of imposition and lifting of the cordon sanitaire as an optimization problem. The gains it
makes in terms of containing the spreading of infection, the costs that need to be paid in terms of higher infections 
within the contained community and the economic fallout due to halting of businesses and finally the constraints that 
the corresponding health infrastructure is able to bear the burden of the growing infections, are the parameters to be 
considered in the problem. We outline here the above mentioned factors from the points of views of (a) early analysis of the data for
COVID-19 and past data of other epidemics, (b) study of compartmentalized models that capture the qualitative picture in 
terms of few parameters, and (c) artificial intelligence (AI) and machine learning (ML) approaches. 

In the early stages of the spreading of COVID-19, data driven approaches were able to trace the correlations of travel patterns and    
infection spreading (see e.g., \cite{chin}) in China. It has a more well documented study for earlier epidemics (SARS \cite{sars}, Ebola \cite{ebola}) although in
a much smaller scale. Nevertheless, it is crucial to study these data driven approaches, given that the effects seen in the real data
for the imposed restrictions were later used as inputs for the various other approaches such as compartmentalized models and also
as training sets for ML approaches. Therefore, in sec. II we outline such studies that essentially correlate the infection spreading with
human traffic. The clear positive correlation in the early stages and a subsequent anti-correlation \cite{covid6} outlines the 
mechanisms of primary and secondary stages of the infections, which are very useful insights for subsequent models. 

The mathematical modeling of epidemic spreading also has a long history \cite{ep1,ep2}. It was the pioneering physicist Daniel Bernoulli who first introduced the mathematical model
approaches of epidemic spreading \cite{bern} in 1766. Since then, the most used model have been the Susceptible-Infected-Removed (SIR) model \cite{sir_model} and
its other variants, generally called compartmentalized models \cite{chen} -- where the total population is divided into groups of populations
and the dynamics of the model proceeds through movements of the populations between these compartments i.e., a susceptible individual
can get infected and then subsequently recover and so on. Extensions of this model include introduction of other plausible compartments
e.g., exposed, representing individuals who came in contact with infected population but not yet showing symptoms. Even further divisions
depending on the severity of the infections can estimate the load of patients needing extensive medical attention. 
The key parameters in these models are the rates at which the populations are relabeled from one compartment to the other i.e., infection rate, recovery rate and
so on. These parameters are often estimated from the data driven approaches mentioned above. Also, the effects of imposed restrictions are
assumed to be mirrored in the variations in these parameters. The models with such estimated parameters and their variations are then used 
to estimate the spreading scale of the epidemic and the possible effects of movement restrictions. Furthermore, given the correlation of the 
scale of epidemic spreading and the negative impact on economy (see e.g., \cite{covid_economy}), it also gives an insight into the economic cost. Therefore, a 
dynamical optimization of the imposed restrictions can be attempted. We outline these efforts in sec. III.

\begin{figure}[tbh]
\centering
\includegraphics[width=17cm, keepaspectratio]{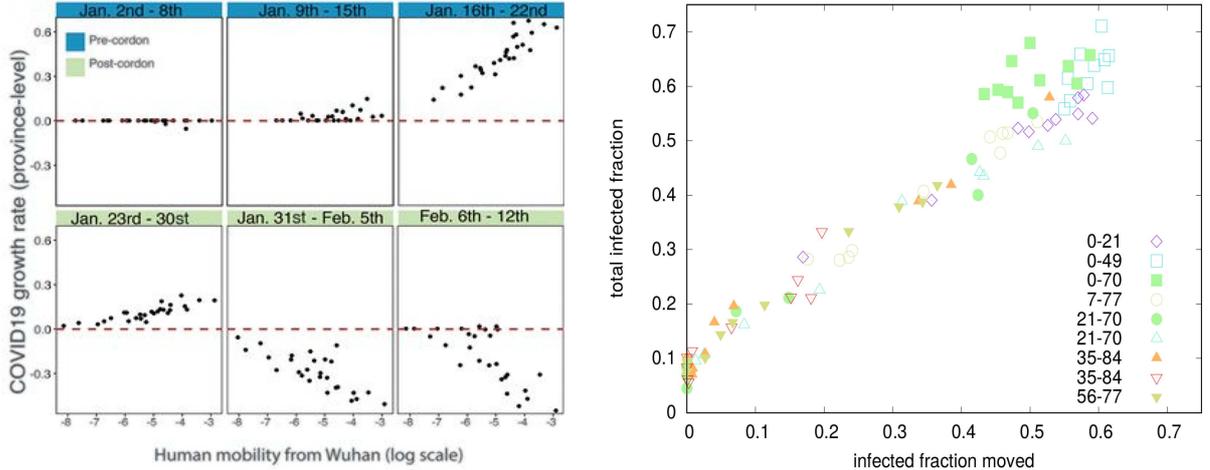} 
\caption{The figure on the left (reproduced from \cite{covid6} with permission), depicts the relationship between the human mobility and the rate of infection in China, before
and after the imposition of cordon sanitaire. There is a clear positive correlation between the two quantities before imposition of restrictions. On the right hand side figure,
simulations of SIR model with optimized mobility of individuals among different regions of varying degree of risks (see \cite{physica}) are shown. For different duration
of travel restrictions (indicated by the start and end dates), the fraction of infected individuals moved correlate strongly with total infection fraction.}
\label{mobility}
\end{figure} 

\begin{figure}[tbh]
\centering
\includegraphics[width=10cm, keepaspectratio]{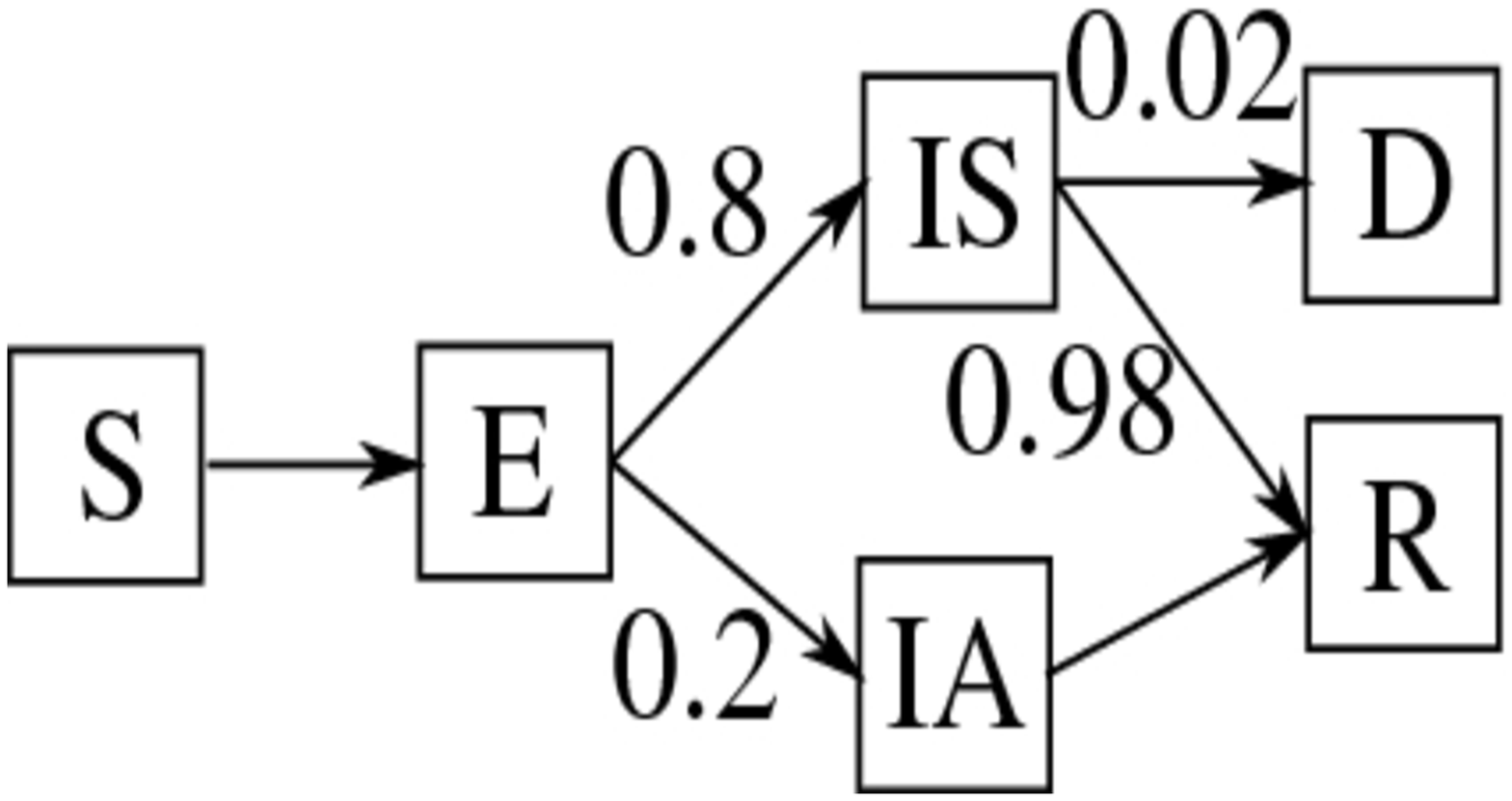} 
\caption{A schematic representation of compartmentalized models of epidemic spreading. The SEIRD model \cite{xing} depicted here
divides the total population, which is assumed to be constant, in different groups and the arrows indicate the directions in which
the population can move from one compartment/state to the other and the numbers indicate the corresponding rates. The numerical
values are estimated depending on the context of the model application (here from Baidu's data) and the mean-field governing equations
are given in Eq. (\ref{seir}).}
\label{schematic}
\end{figure}

Finally, a multidimensional set of data with many attributes is something that can used for a systematic statistical trend analysis
to gain insights that are not immediately apparent. This brings in the machine learning approaches for the study of the real data
for the pandemic. There are specific areas in which the AI-ML approaches can help in advancing our understanding \cite{naude}. The early
warning of the outbreak, the predictions for total infections and/or end-time for the pandemic, implementations of physical
distancing are some such areas. The outstanding challenge in these approaches is the lack of sufficient training data sets that are 
reliable for a stable prediction. In the case of COVID-19, data from previous epidemics (SARS, Ebola, Zika virus) were used in some cases with suitable
adjustments (see e.g., \cite{akhtar,gleam}), in some other cases synthetic data from  optimized parameters of a simplified model \cite{arxiv} 
were also used. The successes and limitations
of these approaches are discussed in sec. IV. 

\section{Data driven approaches to assess effects of travel restrictions}
In drawing any conclusion on the effectiveness of a mitigation strategy for an epidemic, it is essential to analyze 
its effect on the real data. It is often challenging task to have a reliable set of data -- not only due to the 
lack of testing or documentations, but also due to the noise accumulations in news outlets to social media around a
highlighted event \cite{covid_data}. 

Nevertheless, there have been many attempts to understand different aspects of the COVID-19 pandemic, such as 
estimation of reproduction rate \cite{data1}, forecasting of end-time \cite{data2} to effectiveness of protective drugs \cite{data3}, from the analysis of the available data, 

In terms of the movement restriction strategies, at the early stage of the spread of COVID-19, it was possible to trace the correlation between the travel pattern from 
the Hubei province and the detection of infected individuals outside the province. Indeed until end of January, 2020, 
80\% of all cases were detected within the province \cite{covid6} and only after that cases outside the province started rising. 
Kraemer et al. \cite{covid6} studied the human mobility pattern using the data from Baidu Inc. and recorded the effects of 
imposing the cordon sanitaire from 23rd January, 2020. Their finding suggests that the initial bias in the age group and gender
in the detected cases were due to the travel history of those individuals to the Hubei province. Indeed, following the
imposition of the restrictions, those biases eventually disappeared, suggesting that the cases after that time were due to
the secondary infections. Indeed, there was a very clear positive correlation between the COVID-19 growth rate and other provinces 
in China and the human mobility from Wuhan, before the travel restriction was imposed (see Fig. \ref{mobility}). The correlation started decreasing 
after a week of the imposed restrictions and beyond that it showed negative correlation. This implies that an early imposition
of the travel restriction helps in containing the infection, but such restrictions are less useful when the infection has spread
outside a localized region. This was a key observation that formed the basis of the input parameters of the mathematical modeling approaches
that we discuss in the next section.

\section{Compartmentalized models and movement optimization strategies}
There are a myriads of factors that can influence a respiratory infection such as COVID-19. First, the interaction patterns of humans, 
the carriers of the virus, is complex and highly heterogeneous and to a large extent without much of accessible data. Second, 
especially during the first months of the virus spreading, lack of testing facilities contributed to much of the fluctuations in the data. 
Such fluctuations continue even till date, given that a substantial portion of the infected individuals are not symptomatic \cite{zheng}
but can still be infected and thereby can infect others. Third, the effective virulence of the infection is a dynamic quantity. This
is because of the mutation of the virus itself \cite{mutants} and also because of the various restrictive measures imposed. Both of these factors vary 
with time and as well as space. 
 Therefore, the complexity of the system and the noise in the available data
are both very high. 

Nevertheless, attempts to formulate a mathematical model based description of epidemics have been made for over several centuries \cite{bern}. 
This is partly because models provide us with insights that are otherwise inaccessible by simply studying the data. In complex
systems, simplified model approaches have been very useful in gaining critical insight into the system, even though the models
in question ignored many realistic features of the system under study. An outstanding example of success comes from the study of magnetism 
phenomena through the Ising model \cite{ising}. While a drastic simplification over the actual ferromagnetic materials, the model reproduces 
qualitative as well as quantitative features near the paramagnetic to ferromagnetic phase transition point of magnetic materials. 
Even nearly a century after its introduction, the model continues to provide critical insights into the theoretic understanding and practical 
applications for magnetic materials \cite{huang}. 

However, there are a few differences between the analogy of the simplified models for magnetism and that for epidemic spreading. 
First, accurate data are available from experiments in the case of magnetism, which is not true here. While that could still be considered as a noise, 
the second and the most important difference is that in the case of magnetism (and in many other examples of tuned or self-organized criticality) 
the system is near a continuous phase transition. This necessarily imply a scale-free behavior of he system, which broadly means that all fluctuations 
that are of smaller scale than the system size are irrelevant in a renormalization group (RG) sense. This is the precise reason for the universal nature of the
response statistics of magnetic systems near criticality and also the reason why a simplified model devoids of such `small scale' details works 
for such systems near the critical point. Now, in the models of epidemic spreading, there exists no such critical point. Nor do the dynamics
of epidemic spreading are known to produce universal quantitative numbers (similar to, say, the critical exponents) for different instance of epidemics.
Therefore, the arguments of irrelevant parameters in the RG \cite{huang} sense do not hold. 

While the above mentioned criticisms are applicable for epidemic spreading models, there are two points to note before we proceed into the specific modeling approaches. 
First, the goals of epidemic spreading models and that of (laboratory scale) physical systems can vary. With just a model alone, without inputs from real data, no
epidemic model attempts in making quantitative predictions. Second, although a critical point does not exist in epidemic spreading models in itself, 
it has been shown using spatial pattern of the spreading data for COVID-19 pandemic that it follows a fractal growth \cite{covid_fract}. Indeed, it was also 
shown recently \cite{arxiv}
that if a simple model is to make predictions having least errors with the real data, the parameters in the model is to be set in such a way as to have the resulting
spreading pattern in fractal form. Although not arising out of a criticality in the epidemic model, there exist scale free characteristics in such fractal geometry.
With this in mind, we discuss the various models and the results of incorporating movement restrictions in those models for the case of COVID-19 spreading.

\subsection{SIR and related models} 
Bernoulli 
first proposed such an attempt in 1766 \cite{bern}. This class of models are sometimes termed as compartmentalized models, since the basic idea involves
dividing the total population into groups, based on their exposure (or lack of it) to the virus. 
The most used and the most simple version of the model involves dividing (at any instant of time $t$) the entire population into three groups: 
Susceptible $S(t)$, Infected $I(t)$ and Removed $R(t)$.
First proposed in 1927 \cite{sir_model}, the model assumes that the total population $S(t)+I(t)+R(t)=N$ is constant throughout the dynamics. At $t=0$, of course, 
$N=S_0+I_0$, where $S_0$ and $I_0$ represent the initial infection and susceptible population respectively. A susceptible individual, while coming in contact
with an infected individual, can get infected with a certain rate $r$, and an infected individual is removed (due to recovery or death) with a rate $\alpha$.
There is no scope of re-infection in this model, although other variants exist \cite{ep1,ep2} where such scenarios are considered. 

A mean-field treatment of the model is straightforward, which involves writing down the differential equations governing each of the three groups:
\begin{eqnarray}
\frac{dS(t)}{dt} &=& -rI(t)S(t) \nonumber \\
\frac{dI(t)}{dt} &=& rI(t)S(t)-\alpha I(t) \nonumber \\
\frac{dR(t)}{dt} &=& \alpha I(t).
\end{eqnarray}
The temporal evolution of the infected number $I(t)$ from this model in mean-field and in compact lattices behaves in a way similar to a wave of infection in COVID-19
(and other epidemic) spreading. In this form, the model does not give multiple peaks in infections. Indeed, it is easy to see that the maximum value of the
infection will be $I_{max}=I_0+S_0-\frac{1}{q}\left(1+ln(qS_0)\right)$, where $q=r/\alpha=R_0/N$, where $R_0$ is the reproduction rate. The quantity $I_{max}$ 
is important because, this gives the estimate of the 
maximum load the healthcare infrastructure needs to support. SIR models were used in studying effects of optimal migrations (see e.g., \cite{physica}).
 Indeed, in more realistic variants of the model, this is quantity that were estimated for various
different countries in order to make the above mentioned load and also to design optimization strategies of implementing mitigating responses, including travel restrictions.

The above mentioned variant is the simplest one that gives the qualitative features of the current pandemic. However, there are multiple 
other variants of the model that includes more realistic features. These still fall under the category of the compartmentalized models, since the
basic feature of dividing the population into different compartments/states still exists. Among these variants, one is the SEIR model, where the
additional state $E(t)$ denotes the part of the population that are exposed to the virus, but not yet infectious i.e., a finite incubation time
is incorporated in the model (see Fig. \ref{schematic}). This is an important extension, particularly when the maximum case load is to be estimated. 
In Xing et al. \cite{xing}  effects of migrations were explicitly included. The dynamics evolved following the equations:
\begin{eqnarray}
\frac{dS}{dt} &=& -\frac{\beta_1SE+\beta_2SI}{N}+(a-b) \nonumber \\
\frac{dE}{dt}&=&\frac{\beta_1 SE+\beta_2SI}{N}-\delta E+(a-b)E \nonumber \\
\frac{dI}{dt}&=&\delta E-mI+(a-b)I \nonumber \\
\frac{dQ}{dt}&=&mI-\gamma Q \nonumber \\
\frac{dR}{dt}&=&\gamma Q,
\label{seir}
\end{eqnarray}
where $Q$ denotes the confirmed cases, $\delta$ is the infection rate, $\gamma$ is the recovery rate, $m$ is the confirmation rate, $\beta_1, \beta_2$
denote the transmission incidence rates and $a, b$ denote the immigration and emigration rates respectively. The model parameters can then be estimated from actual data
and effects of travel restrictions can be studied.

Another variant of the compartmentalized model is the SIRD mode, where the final state refers to death due to the disease. 
Other than these, there are more case specific variants that, for example, consider various severity of health conditions following
an infection (see e.g., \cite{tuite}). Such details of the model requires additional input from the real data, which are done for some specific countries/regions. 

Apart from adding different states in the original SIR model, another direction of realistic extensions have been to incorporate the effects of the
model topology. The above mentioned mean-field nature of the dynamics can prevail only under well homogeneous mixing of the population, which is
certainly not the case particularly when travel restrictions are imposed. Also, the overly restrictive fixed lattice arrangements, where the 
infections can only spread through nearest or next nearest neighboring individuals, is unrealistic. For both of these limits (lattice models and mean-field),
one way to reach the intermediate realistic scale is to tune the infection rate. Another way to achieve the intermediate state is to modify the
topology in which the model is studied. This can involve pruning the fully connected graph to, say, an Euclidean topology  \cite{ps2}, or to introduce
disorder in the lattice models, say, in terms of site dilution \cite{arxiv}. 

\subsection{Control strategies to reduce population mixing and its early lifting}
As mentioned before, Tuite et al. \cite{tuite} studies a SEIR type model for estimating health
 infrastructure load in Ontario, Canada. The model is structured in 5-year age group layers. 
The interactions within the age groups \cite{moss}, the presence of comorbidities (hypertension, heart diseases, asthma, stroke, diabetes and cancer) were also considered
in estimating the severity of the infection (e.g., required ICU care). The dynamics was initiated with uniformly distributed initial infections and then the
effects of control strategies such as extensive testing and physical distancing measures were studied using a fixed duration and also in a dynamically tuned 
 manner depending on required ICU cares. It was found that dynamically introduced restriction measures were more effective than a fixed duration restriction, with
potentially shorter period of physical distancing. 

In the US also such compartmentalized model (SEAIR) approach was taken to find optimal control in the outbreak \cite{tsay}. The additional state $A(t)$ represents the
estimated 20-40\% of the asymptomatic cases, who can still be carriers of the infection. Here also it was concluded that the effect of interventions (testing, isolation,
physical distancing) are more effective in the early stages of the dynamics than at later stages, even if the measures are more drastic later on. Also, a periodic
on-off strategy, similar to ref. \cite{tuite}, is found to be more effective in controlling the spreading and also conjectured to be more palatable. 

A similar approach was taken by Prem et al. \cite{prem} for the spreading of infection in Wuhan, China. As in Ref. \cite{tsay}, a SEIR model with different age groups having
different rates of infections were studied. The effects of imposing continued restrictions, modeled by taking the corresponding interaction matrices between different age groups, 
seen to lower the total infection rate. Also, an early lifting of such restrictions leads to secondary peaks (see also \cite{physica}).  With a similar SEIR type model, it was shown in ref. \cite{kuc}
that the effective reproduction index $R_t$ decayed 2.35 on January 16 (one week before cordon sanitaire) to 1.05 on January 31 (one week after cordon sanitaire). This also
reinforce the benefit of early imposition of restrictions. 

\subsection{Cordon sanitaire as an optimization problem}
As discussed above, there is a general consensus regarding the benefit of early imposition of cordon sanitaire in reducing the load on healthcare systems. A subsequent dynamical 
(on-off) interventions (travel restrictions), rather than a prolonged period of restriction, also seem to work better in reducing the total spreading. However, the optimization 
needs to consider the relative rates in which the cordoned-off and the remaining population gets affected. Also, while relaxing the restrictions, the optimization function
for an individual may not be the same as the global optimized state.

\begin{figure}[tbh]
\centering
\includegraphics[width=15cm, keepaspectratio]{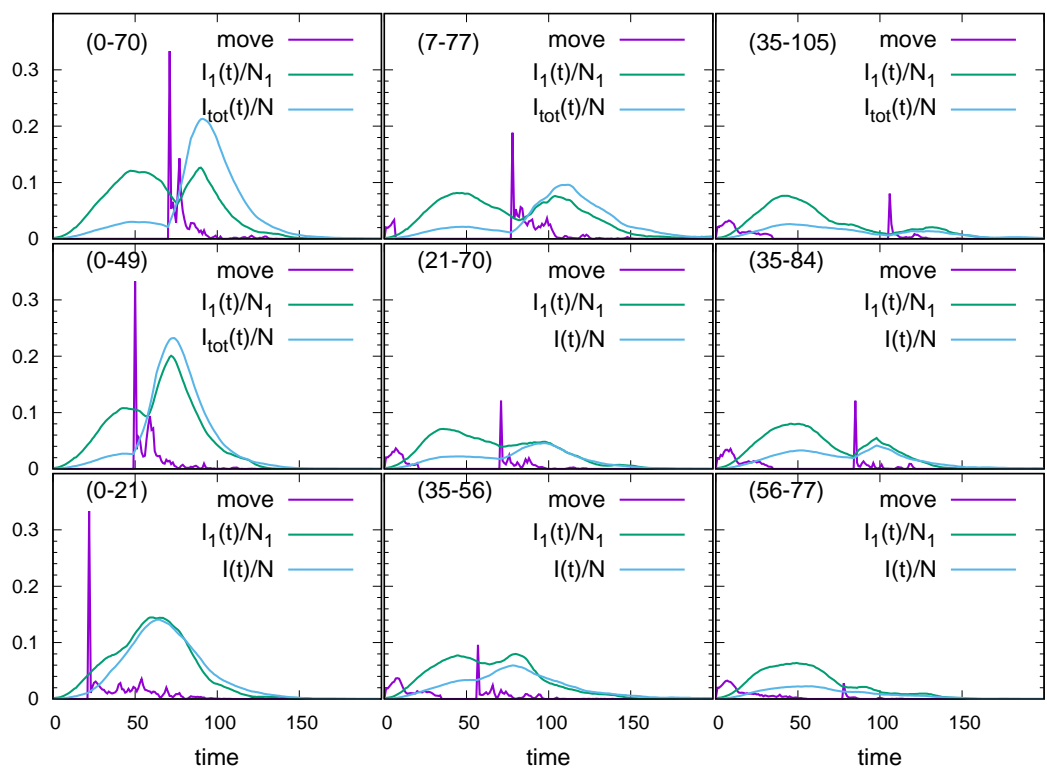} 
\caption{Simulations of SIR model with optimized movements between communities of 
different risks (reproduced with permission from \cite{physica}). It is seen that the 
infected fraction in the high-risk region $I_i(t)/N$ shows a secondary peak once the travel
restrictions are lifted early (time period indicated in the figures). However, the overall
(relative) size of the second-wave peak ($I_{tot}/N$) is seen to be larger (see also \cite{esp,prem}). }
\label{second_wave}
\end{figure} 

Espinozo et al. \cite{esp} noted that when unrestricted movements are allowed between two low risk communities, the chances of secondary infections increase in those communities, but the overall epidemic size is reduced. On the other hand, imposition of cordon sanitaire around a high risk community -- the original practice of such type -- reduces secondary infections, but increases the overall epidemic size, since the infection greatly affects the high risk community. Therefore, it is not straightforward to asses the benefits of such travel restrictions and also the time of removal of such restrictions. Indeed, the overall process of implementing the mitigation strategies can be viewed from the point of view of control theory \cite{stefan} with a limit on the maximum active cases as a constraint that represent the load on health care infrastructure.  In the following section, we will discuss whether a machine learning approach can optimize the restriction times, so as to limit the spreading of epidemic. Before that, however, it is also interesting to note that while the objective of the governments would be to optimize the travel restrictions so as to minimize the epidemic size at a reduced economic fallout, from the point of view of an individual, that objective may not match. Particularly, given a chance, an individual would travel to a lower risk community rather than to stay in a higher risk community. But given that many other individuals might also try the same, the said low risk community might not remain low risk due to spreading of secondary infections. This situation can be viewed from a game theory perspective \cite{physica}, where the situation is that of a set of coupled minority games, played in parallel. It was seen that a restriction on the number of travel upon an individual is more effective than imposing a full stoppage of travel. But similar to what is noted in ref. \cite{prem}, for example, an early lifting of the restrictions can bring a second wave of infections (Fig. \ref{second_wave}).

\section{Machine learning approaches}
Here we aim to revisit the recent scientific contributions based on Artificial Intelligence (AI) to the fight against COVID-19 pandemic. In recent past applications of AI in different aspects of epidemiology is instrumental in policy and medical analysis measuring the cost of the pandemic in terms of lives and economic damage (see e.g., \cite{ai_covid1,ai_covid2}) etc. The recent literature ranges from early warning, tracking and prediction to social control which often influence the migration of the people to avoid the viral disease (see e.g., \cite{naude}).  
In January 2020 China imposed very strict lockdown to contain the very first Covid-19 outbreak, which were in place till April 2020. During that period researchers were speculative about the impact of these policies on virus spreading.  The AI based techniques are primarily used to predict the duration of social restrictions in different geographical regions as it helped in reducing the number of infections significantly.   In this direction Yang et al. \cite{yang} employed a modified susceptible-exposed-infected-removed (SEIR) epidemiological strategy to predict the epidemic progression by including the people’s migration data prior to and after the January 2020 along with the COVID-19 epidemiological at that point of time.  The authors used the Long-Short-Term-Memory (LSTM) model of recurrent neural network (RNN) to estimate the number of newly infected people by processing various time-series problems. The 2003 SARS outbreak statistics is used to train the devised model. The devised model fed with the COVID-19 spreading parameters, such as the rate of spreading, infection probability, recovery rate etc. This SEIR based approach was useful in estimating peaks and sizes of the COVID-19 epidemic. The model constrained by the inadequate data set which results in relatively simple network configuration and may suffer from overfitting problem.
In a similar study Xing et al. \cite{xing} studied the impact of migration of people using Baidu’s migration data of Guangdong and Hunan provinces. As mentioned before, the author developed a three-stage dynamical model. It uses SIER, where a time variant function is used for susceptible S(t), infected I(t), exposed E(t) and removed R(t) individuals (see Fig. \ref{schematic}). In the first stage i.e., early stage of the epidemic spreading the model assumes that the confirmed individuals Q(t) are not migrating. And the COVID-19 transmission dynamical modeling is represent using Eq. (\ref{seir}). The model parameters were estimated using mobility data from Baidu.

 Further, very similar models were used for the remaining two stages to characterize the imposition of the social curbs and resumption of the regular life respectively.  Afterwards the mathematical analysis of only first stage is carried out and reproduction rate  is calculated. The other parameters values were calculated from Baidu’s data and using the methodologies such as least-square method. The result shows that scale of infection is low in the province which emigrated the population. However, the province receives the population is exactly the opposite. And the authors predicted that the province which emigrated the population in the first stage may suffer after the easing of the social curbs (see also \cite{physica} in this context). However, this work suffers from many shortcomings such as limited and erroneous data availability, not considering the asymptomatic population and spatial diffusion characteristics.

To study the health and economic impact Khadilkar et al. \cite{khad} devised an AI-based system. It predicts the best possible lockdown policies to control Covid19 spreading and minimize its economic impact. The reinforcement learning based approach learns from different policies which are represented as a function of disease and population parameters. The disease progression model is primarily based on SIER as depicted in the Fig. \ref{schematic}. Where $S$ is susceptible, $E$ represents exposed, $IS$ represents infected, $IA$ indicates asymptomatic, $D$ is dead, and $R$ indicates recovered individuals. Further the number indicates the probabilities in the transitions from one state to another. The The approach exposes the limitations of the imperfect lockdowns and it can be utilized to investigate various policies by using tunable parameters. Further, the model may be useful to determine more fine-grained social curbs to prevent the COVID-19 spreading.

In another reinforcement learning based approach by Ohi et al. \cite{ohi} demonstrated how an agent’s actions may have different possible outcome based on the spreading of the disease and economic conditions. A virtual pandemic is similar to the COVID-19 is simulated to train the system. Afterwards the training the agent chooses the optimal strategy which reduces epidemic spreading in a financially viable manner. The analysis of the results shows that, to reduce the first surge of infections the system opted for a longer period of lockdown. Again, to curb the successive waves of infections the system chooses a combination of recurrent lockdowns and shorter periods of lockdowns. Although, the model is able to provide a middle ground between epidemic spreading and economic gains. However, a comparative study between humanitarian loss and economic gains when total lockdown is imposed and when recurrent lockdown would have been interesting.

\section{Discussions}
The first response of the governments in most of the countries to the outbreak of COVID-19 have been to impose restrictions on human mobility from high infected regions. 
Following the spreading of the virus in most countries in the world, the subsequent response have been to quarantine the infected population and also to impose local restrictions on human mobility. Those restrictions, while helpful in limiting the maximum number of active cases, did have and will continue to have severe societal and economic impacts. Here we reviewed the multiple facets of such restrictions on mobility in different countries, based on the analysis of data, study of models and machine learning approaches. The emerging picture seems to be that while an early imposition of restrictions are useful, for the subsequent period, a periodic relaxations of the restrictions is a more effective strategy than to have a prolonged imposition of restrictions. 

The process of optimizing the restriction period is not straightforward and likely to differ among different countries, based on their socio-economic activities and healthcare infrastructure. A major challenge in finding such optimized strategy has been to gather noise-free data regarding the spreading dynamics of the virus. Due to the complex nature of the human interactions, compartmentalized modeling approaches are also hard to implement. However, we have discussed various efforts that address these issues. For example, an age-based hierarchy in the models seem to help the optimization process, given that the nature of interactions and required health-care vary among different age groups. Also, in using data driven machine learning approaches, use of earlier epidemic data for training can be a useful strategy.


\end{document}